# Mutual Coupling of two Semiconductor Quantum Dots via an Optical Nanocavity

A. Laucht,[1] J. M. Villas-Bôas,[1,2] S. Stobbe,[3] N. Hauke,[1] F. Hofbauer,[1]
G. Böhm,[1] P. Lodahl,[3] M.-C. Amann,[1] M. Kaniber,[1] and J. J. Finley[1,*]

[1]*Walter Schottky Institut, Technische Universität München, Am Coulombwall 3, 85748 Garching, Germany*
[2]*Instituto de Física, Universidade Federal de Uberlândia, 38400-902 Uberlândia, MG, Brazil*
[3]*DTU Fotonik, Department of Photonics Engineering,
Technical University of Denmark, Ørsteds Plads 343, DK-2800 Kgs. Lyngby, Denmark*
(Dated: July 12, 2010)

We present an experimental and theoretical study of a system consisting of two *spatially separated* self-assembled InGaAs quantum dots strongly coupled to a single optical nanocavity mode. Due to their different size and compositional profiles, the two quantum dots exhibit markedly different DC Stark shifts. This allows us to tune them into mutual resonance with each other and a photonic crystal nanocavity mode as a bias voltage is varied. Photoluminescence measurements show a characteristic triple peak during the double anticrossing, which is a clear signature of a coherently coupled system of three quantum states. We fit the entire set of emission spectra of the coupled system to theory and are able to investigate the coupling between the two quantum dots via the cavity mode, and the coupling between the two quantum dots when they are detuned from the cavity mode. We suggest that the resulting quantum V-system may be advantageous since dephasing due to incoherent losses from the cavity mode can be avoided.

PACS numbers: 42.50.Ct, 42.70.Qs, 71.36.+c, 78.67.Hc, 78.47.-p

## I. INTRODUCTION

Cavity quantum electrodynamics experiments (cQED) using semiconductor quantum dots (QDs) have recently attracted much interest in the solid-state quantum optics community.[1,2] Much progress has been made with a number of spectacular demonstrations, including efficient generation of non-classical light,[3] the observation and investigation of strong coupling phenomena[4–8] and, excitingly, the possibility to observe and exploit quantum optical non-linearities.[9,10] These developments are all ingredients for the realization of a solid-state all-optical quantum network[11] where quantum memory elements are coupled via single light quanta. In 1999 Imamoğlu *et al.*[12] proposed that two spatially separated electron spins in QDs could be coherently coupled via a common optical cavity field. However, it was only in the past five years that the strong coupling regime was reached for a single QD[4–6] and, to the best of our knowledge, only one observation of two dots coherently interacting with a common cavity mode has been published.[7] This would provide a new way to entangle spatially separated quantum emitters via the electromagnetic quantum vacuum.

In this paper, we present experimental and theoretical investigations of a system of two *spatially separated* QDs strongly coupled to the same high-Q photonic crystal (PC) nanocavity mode. We identify the two different QDs via their different voltage dependent shifts when we tune their emission energies via the quantum confined Stark effect (QCSE).[13–15] Furthermore, we use the same effect to tune their emission energies into resonance with each other and through resonance with a cavity mode, which is energetically close to their crossing point. The photoluminescence (PL) data shows a characteristic triple peak during the double anticrossing which is an unambiguous signature for the coherent coupling of the three quantum states.[33] Previously, the authors of Ref.[7] presented a double dot micropillar system operating in the strong coupling regime but did not analyze the spectral function of the system and its dependence on detuning. Here, we obtain new information by theoretically modelling the spectral function of our system and fitting the experimental data using the model introduced in[8,16], extended to include two different quantum emitters. By fitting this model to our data we extract the contributions of each quantum state to the three branches of the double anti-crossing in the spectral emission. This comparison clearly indicates that coherent coupling is established between the two separated QDs, and, moreover allows us to identify a situation where the two QD states have the same detuning from the cavity mode. The strongly coupled double dot - cavity system then forms a V-like three level quantum system[17] (see Fig. 1 (b) (inset)) where entanglement between the two different QD excitons is established by virtual photon emission and absorption, without populating the cavity mode. This would allow to circumvent the dominant source of decoherence in solid-state cQED, photon loss from the cavity. Since it is technologically difficult to further enhance the Q-factor of GaAs based photonic crystal nanocavities, this would provide a route towards creating cavity mediated entangelement of the two QD excitons on the basis of currently existing solid-state technology. In this case photon loss from the cavity, which determines the mode Q-factor, is no longer important for the cavity mediated coupling of the two QDs.

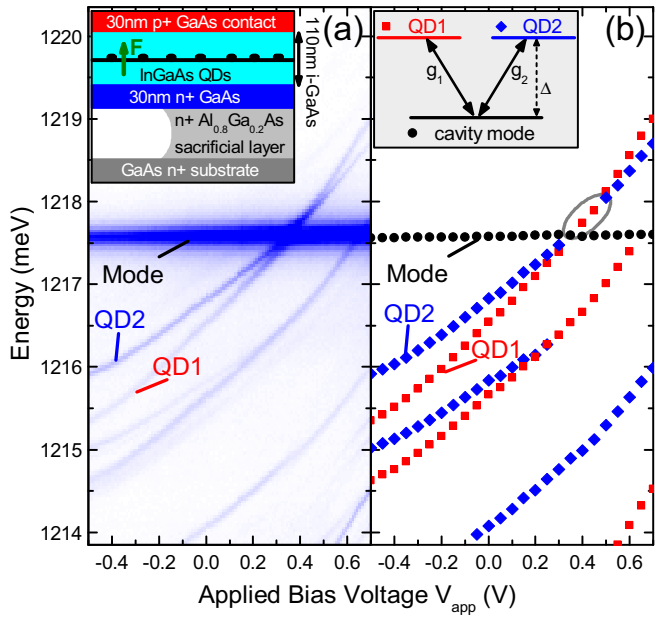

FIG. 1: (color online) (a) PL spectra of the double dot-cavity system as a function of bias voltage (false-colour plot). (inset) Schematic of the layer structure of the device. (b) Peak positions of cavity mode (black circles), QD1 (red squares) and QD2 (blue diamonds) for different bias voltages, emphasizing the different DC Stark shifts of QD1 and QD2. (inset) Schematic of the energy levels when QD1 and QD2 are in resonance and both are detuned from the cavity mode, forming a V-system (indicated by the gray ellipse at $V_{app} = 0.45$ V).

## II. SAMPLE STRUCTURE AND EXPERIMENT

The sample investigated was a GaAs $p$-$i$-$n$ photodiode with PC defect nanocavities patterned into a free standing membrane.[15,18] The layer sequence is depicted schematically in the inset of Fig. 1 (a). Low density InGaAs self-assembled QDs ($\leq 20$ $\mu m^{-2}$) were incorporated into the middle of the i-region of the device, allowing vertical electric fields to be applied by varying the voltage across the junction ($V_{app}$). Electrical contacts were established to the $n-$ and $p$-doped contact layers using optical lithography and 250 $\mu m \times 400$ $\mu m$ photodiode mesas were formed using wet chemical etching.[18] A modified $L3$ defect[19] at the center of each PC is used to generate a strongly localized, high-Q nanocavity. Such cavities support six modes within the 2D-photonic bandgap of which the fundamental cavity mode exhibits Q-factors up to $\sim 12000$ in our structures, high enough to reach the strong coupling regime.[5,8] PL spectra from the nanocavities were recorded using confocal microscopy as a function of $V_{app}$. The excitation laser was tuned to the second order higher energy cavity mode in order to spatially select only dots coupled to the fundamental mode.[20,21]

In Fig. 1 (a) we present PL data recorded as a function of $V_{app}$ as an overview of the investigated situation. Besides the emission of the fundamental cavity mode that is not influenced by $V_{app}$, we observe a number of QD emission lines that clearly shift as $V_{app}$ varies, due to the QCSE.[15] Two classes of lines are observed, each exhibiting a markedly different voltage dependence. We attribute the two classes of lines to two different QDs with different size and In-Ga compositional profiles. This leads to intrinsically different distributions of the electron and hole wavefunctions[13,22] and, consequently, different polarizabilities of the exciton transitions. As a result the different QD exciton lines can be electrically tuned relative to one another. Fig. 1 (b) shows the extracted peak positions of all lines shown in Fig. 1 (a) for better comparison. The black circles, red squares and blue diamonds correspond to emission from the cavity mode, QD1 and QD2, respectively. Two of these exciton lines from different QDs (labelled QD1 and QD2) cross at an energy of $\hbar\omega_X = 1217.8 \pm 0.1$ meV, very close to the energy of the cavity mode ($\hbar\omega_c = 1217.6 \pm 0.1$ meV). Both these transitions can be assigned to single excitons due to the observed linear power dependence of their emission intensity (data not shown). By changing $V_{app}$, the QD1 and QD2 excitonic transitions can be shifted through the cavity mode and are resonant at almost the same voltage. We note that at this value of $V_{app}$ the carrier tunnelling escape rate from the dots is negligible and the emission intensity is field independent.[15]

The system described above allows us to investigate the cavity mediated coupling between QD1 and QD2 in two different ways. Depending on $V_{app}$, the states of the dots can be simultaneously tuned into resonance with the cavity mode where coherent coupling occurs via the cavity field. Alternatively, they can be in resonance with each other but detuned from the cavity mode to form a degenerate V-system (see Fig. 1 (b) - inset). Most importantly, in the latter case dephasing due to incoherent loss of photons from the cavity, the dominant source of dephasing, can be circumvented as discussed below.

## III. THEORY

In order to describe the system theoretically we extend our previously presented model for a single QD exciton[8] to include two independent excitons coupled to a common cavity mode. We use the following Hamiltonian:

$$H = \sum_{n=1}^{2}\left[\frac{\hbar\omega_n}{2}\sigma_z^n + \hbar g_n(a^\dagger\sigma_-^n + \sigma_+^n a)\right] + \hbar\omega_c a^\dagger a \quad (1)$$

where $\sigma_+^n$, $\sigma_-^n$ and $\sigma_z^n$ are the pseudospin operators for the two level system consisting of ground state $|0\rangle$ and a single exciton $|X_n\rangle$ state of the $n^{th}$–QD $\{n = 1, 2\}$. $\omega_n$ is the exciton frequency, $a^\dagger$ and $a$ are the creation and destruction operators of photons in the cavity mode with frequency $\omega_c$, and $g_n$ describes the strength of the dipole coupling between cavity mode and exciton of the $n^{th}$–QD. The incoherent loss and gain (pumping) of the dot-cavity system is included in the master equation of the Lindblad form

$$\frac{d\rho}{dt} = -\frac{i}{\hbar}[H, \rho] + \mathcal{L}(\rho), \quad (2)$$

where

$$\begin{aligned}\mathcal{L}(\rho) &= \sum_{n=1}^{2} \Big[\frac{\Gamma_n}{2}(2\sigma_-^n\rho\sigma_+^n - \sigma_+^n\sigma_-^n\rho - \rho\sigma_+^n\sigma_-^n) \\ &+ \frac{P_n}{2}(2\sigma_+^n\rho\sigma_-^n - \sigma_-^n\sigma_+^n\rho - \rho\sigma_-^n\sigma_+^n) \\ &+ \frac{\gamma_n^\phi}{2}(\sigma_z^n\rho\sigma_z^n - \rho)\Big] + \frac{\Gamma_c}{2}(2a\rho a^\dagger - a^\dagger a\rho - \rho a^\dagger a) \\ &+ \frac{P_c}{2}(2a^\dagger \rho a - aa^\dagger \rho - \rho aa^\dagger).\end{aligned}$$

Here, $\Gamma_n$ is the exciton decay rate, $P_n$ is the rate at which excitons are created by a continuous wave pump laser, $\gamma_n^\phi$ is the pure dephasing rate of the exciton in the $n^{th}$-QD, which accounts for effects originating from high excitation powers or high temperatures, $\Gamma_c$ is the cavity loss and $P_c$ is the incoherent pumping of the cavity.[34]

Assuming that most of the light escapes the system through the radiation pattern of the cavity and using the Wiener-Khintchine theorem, the spectral function is then given by[17]

$$S(\omega) \propto \lim_{t\to\infty} \text{Re} \int_0^\infty d\tau e^{-(\Gamma_r - i\omega)\tau} \langle a^\dagger(t) a(t+\tau)\rangle, \quad (3)$$

where the term $\hbar\Gamma_r = 18$ μeV (half-width) was added to take into account the finite spectral resolution of our double-monochromator.[23] The emission eigenfrequency is obtained by solving the Liouvillian equations for the single time expectation value, similar to ref.[24]:

$$i\frac{\partial}{\partial t}\begin{pmatrix} \langle a \rangle \\ \langle \sigma_-^1 \rangle \\ \langle \sigma_-^2 \rangle \end{pmatrix} = \begin{pmatrix} \bar{\omega}_c & g_1 & g_2 \\ g_1 & \bar{\omega}_1 & 0 \\ g_2 & 0 & \bar{\omega}_2 \end{pmatrix} \begin{pmatrix} \langle a \rangle \\ \langle \sigma_-^1 \rangle \\ \langle \sigma_-^2 \rangle \end{pmatrix} \quad (4)$$

where $\bar{\omega}_c = \omega_c - i\gamma_c$ and $\bar{\omega}_n = \omega_n - i\gamma_n$, with $\gamma_c = (\Gamma_c - P_c)/2$, and $\gamma_n = \gamma_n^\phi + (\Gamma_n + P_n)/2$. From the eigenstate of the emission eigenfrequency we can obtain the degree of mixture of each peak in the spectrum, i.e. the strength of the contribution of cavity mode, QD1 exciton and QD2 exciton to each individual eigenstate.

## IV. RESULTS

We investigate the described system experimentally and theoretically in Fig. 2. Panel (a) shows high-resolution PL spectra plotted in a false-color plot recorded as a function of $V_\text{app}$. Whilst the cavity mode is unaffected by the electric field, the two QD excitons shift into resonance with the mode at $V_\text{app} \sim 0.4$ V via the QCSE and into resonance with each other at $V_\text{app} \sim 0.5$ V. During the double anticrossing, when both excitons and the cavity mode are tuned into mutual resonance, we observe a triple peak feature which is an unambiguous sign for the coherent coupling of three quantum states. Using the spectral function $S(\omega)$ we globally fit the *entire set* of spectra generated as a function of $V_{app}$ using a Levenberg-Marquardt algorithm in the same manner as reported in Ref.[8]. The best fit was obtained for $\hbar g_1 = 44$ μeV and $\hbar g_2 = 51$ μeV, $\hbar\Gamma_{QD1} = 0.1$ μeV and $\hbar\Gamma_{QD2} = 0.8$ μeV, $\hbar P_{QD1} = 1.5$ μeV and $\hbar P_{QD2} = 1.9$ μeV, $\hbar\gamma_{QD1}^\phi = 20$ μeV and $\hbar\gamma_{QD2}^\phi = 9.8$ μeV, $\hbar\Gamma_c = 147$ μeV, and $\hbar P_c = 5.7$ μeV.

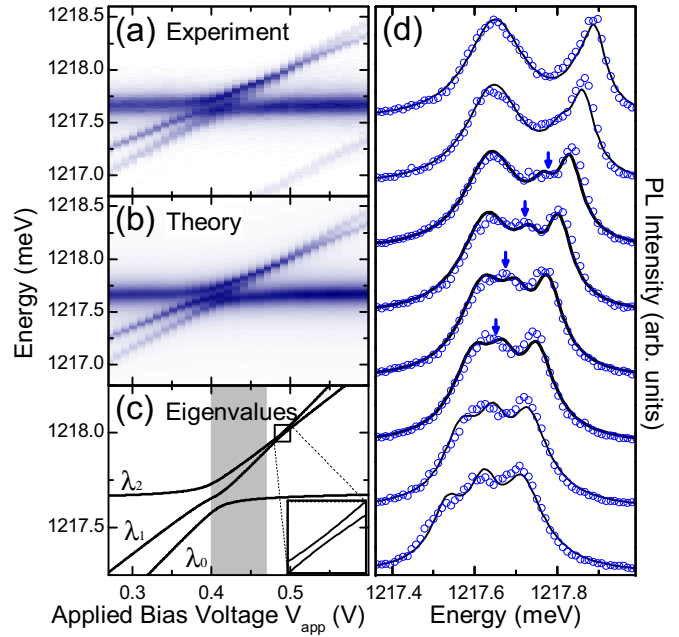

FIG. 2: (color online) (a) High-resolution PL spectra of the investigated system as function of $V_\text{app}$ (false-colour plot). (b) Calculated spectral function of the same system. Parameters were obtained from best-fits. (c) Calculated eigenvalues $\lambda_0$, $\lambda_1$ and $\lambda_2$ (solid black lines), clearly showing the three branches of the double anticrossing and the triple peak in resonance. The predicted anticrossing of the two QDs when they are detuned from the mode is magnified in the inset. For the gray shaded area, we compare in (d) the observed (open circles) and the fitted, theoretical (black solid lines) spectral functions, showing the good agreement.

The calculated spectral function presented in Fig. 2 (b), is showing very good quantitative agreement with the measured data. The spectra close to resonance are plotted in Fig. 2 (d) to directly compare the experimental data (open symbols) and the calculated spectral function (black solid lines). The blue arrows mark the position of the third (middle) peak in resonance which remains visible for all detunings, and is very well reproduced by our calculations with three coupled quantum states. The comparison with the experimental spectrum, which shows the triple peak structure, strongly supports our conclusion that we observe two excitons from two independent QDs. In Fig. 2 (c) we plot the calculated eigenvalues of the matrix in Eq. 4 (i.e. the peak positions) $\lambda_0$, $\lambda_1$ and $\lambda_2$ (black curves). The data clearly indicate the exact evolution of the three branches as function of $V_\text{app}$. Besides the double anticrossing in resonance with the cavity mode, our calculations indicate that an anticrossing must exist for the two QDs when they are tuned into exact resonance with one other, but not with the mode (see magnified inset). Here, the splitting was numerically determined to be ∼ 10 μeV. This energy is below the spectral resolution of our setup and cannot, therefore, be directly resolved by our experiments.[35] Finally, we note that we have observed similar double anticrossings in two other samples (data not shown).

## V. DISCUSSION

The good agreement between experiment and theory (c.f. Fig. 2) lends support to our model and allows us to extract additional information about the coupling between the cavity mode and the excitons. The calculation of the eigenvectors for each of the eigenstates reveals the contributions of cavity mode, QD1 and QD2 to the three different branches of our system. We plot the results in Fig. 3, where the upper, middle and bottom panels correspond to the eigenstates $\lambda_0$, $\lambda_1$ and $\lambda_2$, respectively. The different curves correspond to the normalized admixture of QD1 (red dotted), QD2 (blue dashed) and cavity mode (green solid) to the quantum states of the coupled system. Starting with $\lambda_0$ (upper panel) we can trace the curves for increasing $V_{app}$ to monitor the evolution from an almost pure QD1-like state to the mode-like state with only a weak contribution of QD2 close to resonance. The eigenstate $\lambda_1$ (middle panel) is initially QD2-like, becomes a mixture of all three states for $V_{app} \sim 0.4$ V, is mainly QD1-like for $V_{app} \sim 0.45$ V, and becomes a mixture of QD1 and QD2 only for $V_{app} \sim 0.49$ V. For large $V_{app}$ it remains QD2-like since the system is strongly detuned. For this eigenvalue we can nicely see where a coupling between QD1 and QD2 can occur. When both excitons are in resonance with the cavity mode we obtain a coupling between the two excitons, but the coupling strength is governend by incoherent losses of the mode which effectively limits the coherent interaction. However, when QD1 and QD2 are tuned into resonance with each other, but not in resonance with the cavity mode ($V_{app} \sim 0.49$ V), the system can be seen as a V-type system as depicted schematically in Fig. 1 (b) - inset. The coupling occurs via a Raman-type transition as proposed by Imamoğlu et al.[12]. Here, incoherent losses of the cavity mode, that dominate the dissipation, are only of minor importance. This increases the coherence of the system and relaxes the stringent criteria demanding high Q-factors of the cavity mode, needed for the coupling of the two QDs. The last state $\lambda_2$ (bottom panel) starts mode-like, becomes a strongly mixed state of mode and QD2 for $V_{app} \sim 0.4$ V, is QD2-like for $V_{app} \sim 0.45$ V, and becomes a mixture of QD1 and QD2 for $V_{app} \sim 0.49$ V before it ends QD1-like. For this state, we can also see the coupling of QD1 and QD2 when they are not in resonance with the mode, due to the mixed character.

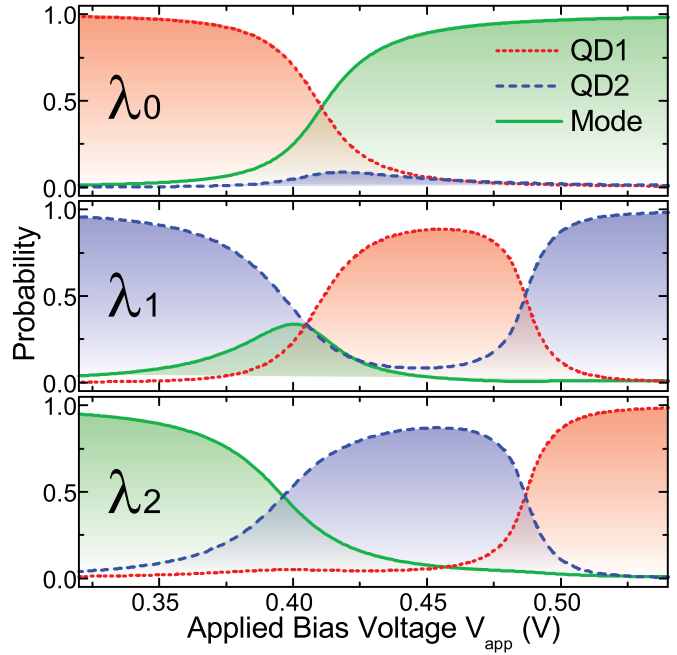

FIG. 3: (color online) Calculated eigenvectors of the investigated system for the eigenvalues $\lambda_0$ (upper panel), $\lambda_1$ (middle panel) and $\lambda_2$ (bottom panel). The plotted curves show the contributions of QD1 (red dotted lines), QD2 (blue dashed lines), and the cavity mode (green solid lines) to the individual states for different detunings as a function of $V_{app}$.

## VI. CONCLUSION

We presented an experimental and theoretical study of a system where two QDs are coherently coupled via an optical cavity mode. Coupling was established by electrically tuning both QDs into mutual resonance and into resonance with the cavity mode, or by tuning them in resonance with each other but detuned from the mode. We pointed out that the latter configuration offers the advantage that photon loss from the cavity can be circumvented, leaving the system in a state of coherent superposition for a longer time and relaxing the stringent criteria for having extremely high mode Q-factors.

## VII. ACKNOWLEDGEMENTS

We gratefully acknowledge financial support of the DFG via the SFB 631, Teilprojekt B3, the German Excellence Initiative via NIM, and the EU-FP7 via SOLID. JMVB acknowledges the support of the Alexander von Humboldt Foundation, CAPES, CNPq and FAPEMIG.

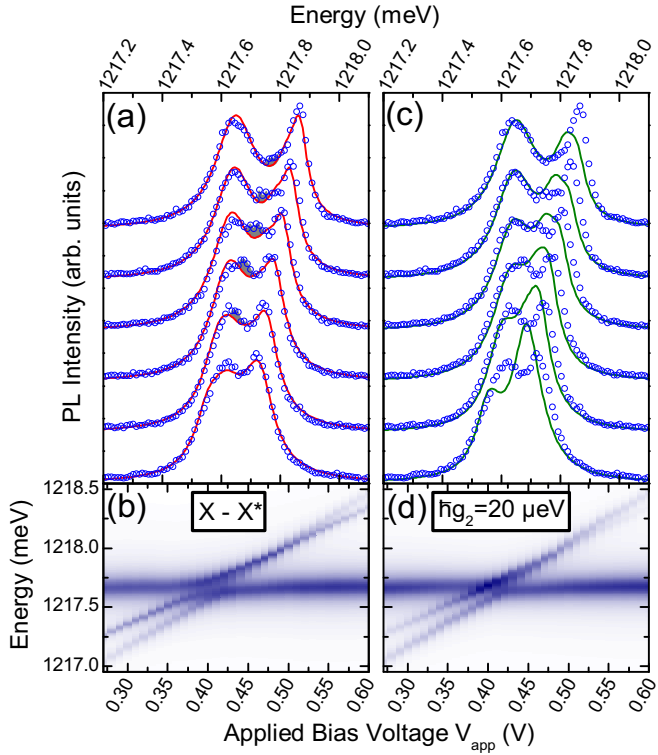

FIG. 4: (color online) Resonant spectra ($V_{app} = 0.41 - 0.46$ V) obtained from fitting of an alternative theoretical model to the experimental data (blue circles), assuming that (a) the two quantum dot states cannot coexist (e.g. neutral exciton and charged exciton of the same quantum dot), and (c) QD2 is only weakly coupled to the cavity mode with a fixed $\hbar g_2 = 20$ μeV. (b) and (d) Contour plots (false color) of the calculated spectra of the same data for a wider range showing the qualitative evolution of the peaks during the crossing.

## VIII. APPENDIX: ALTERNATIVE INTERPRETATIONS

In this section we investigate the possibility of ambiguity in the interpretation of our data. We investigate the possibility that the observed spectral signature could be produced by (i) two different single exciton transitions of the same quantum dot, and (ii) by two different QDs, one weakly and one strongly coupled to the cavity. In Fig. 4, we compare the experimental data with calculations obtained for these two different, alternative models.

In Fig. 4 (a) we present spectra obtained from fitting the experimental data (blue circles) using the same fitting procedure as in section IV, but assuming that the states QD1 and QD2 cannot coexist at the same time (red solid lines). This would be the case for two different states of the same QD, e.g. exciton and charged exciton. While the overall agreement between the calculated spectra and the experimental data is good,[36] the agreement is less good close to resonance. In this situation the resulting spectral function does not exhibit a triple peak, rather it is the sum of the spectra of two independent quantum systems[8] $S(\omega) = (S_1(\omega) + S_2(\omega))$ and, therefore, shows only a double peak close to resonance. We mark the significant difference between the experimental data and the calculated spectral functions with the gray shaded area on the figure. Fig. 4 (b) shows the same calculations in a contour plot presentation. In this plot, the third peak in resonance is, of course, also missing. This observation and the difference in the observed Stark shifts (c.f. section II) excludes this possible explanation.

In Fig. 4 (c) we present calculations obtained, employing the very same model as in section IV, but assuming that QD2 is only weakly coupled to the cavity mode with a fixed $\hbar g_2 = 20$ μeV (green solid lines), and compare it with the experimental data (blue circles). The quantitative agreement between calculations and experiment is clearly unsatisfactory. The qualitative agreement can be evaluated from the contour plot in Fig. 4 (d). Since QD2 is only weakly coupled to the cavity mode, no exciton polaritons form and its spectral position is not affected by the presence of the cavity mode. It simply crosses the cavity mode on the path dictated by the quantum confined Stark shift. Furthermore, in resonance with the cavity mode the emission of QD2 is enhanced due to the Purcell effect, which is a clear signature of weak coupling. However, this situation does not show any similarity to the spectral function observed in the experiments and plotted in Fig. 2 (a).

Of the three presented models, only the genuine model employed in the main text is able to satisfactorily fit the experimental data and represent all features observed in the spectral function of the system. We, therefore, conclude that the QD - cavity system investigated in our experiments, consists of two independent QDs which are both strongly coupled to one common cavity mode.